%% file: SMP-13-015_temp.tex
\pdfoutput=1

\documentclass[11pt,twoside,a4paper,cmspaper,final,collab]{cms-tdr}
\def\svnVersion{275971}\def\cmsCernNo{2013-037}\def\cmsCernDate{\today}\def\cmsMessage{Published in Physical Review Letters as \href{http://dx.doi.org/10.1103/PhysRevLett.114.051801}{\doi{10.1103/PhysRevLett.114.051801.}}}

\begin{document}\cmsNoteHeader{SMP-13-015}

\hyphenation{had-ron-i-za-tion}
\hyphenation{cal-or-i-me-ter}
\hyphenation{de-vices}
\RCS$HeadURL: svn+ssh://svn.cern.ch/reps/tdr2/papers/SMP-13-015/trunk/SMP-13-015.tex $
\RCS$Id: SMP-13-015.tex 273010 2015-01-06 19:38:29Z klute $
\newcommand{\mll}{\ensuremath{m_{\ell\ell}}\xspace}
\newcommand{\ptlmax}{\ensuremath{\pt^{\ell,\text{max}}}\xspace}
\newcommand{\dyll}{\ensuremath{\cPZ/\gamma^*\to\ell^+\ell^-}}
\newcommand{\V}{\ensuremath{\cmsSymbolFace{V}}\xspace}

\newlength\cmsFigWidth
\ifthenelse{\boolean{cms@external}}{\setlength\cmsFigWidth{0.85\columnwidth}}{\setlength\cmsFigWidth{0.4\textwidth}}
\ifthenelse{\boolean{cms@external}}{\providecommand{\cmsLeft}{top}}{\providecommand{\cmsLeft}{left}}
\ifthenelse{\boolean{cms@external}}{\providecommand{\cmsRight}{bottom}}{\providecommand{\cmsRight}{right}}
\ifthenelse{\boolean{cms@external}}{\providecommand{\CL}{C.L.\xspace}}{\providecommand{\CL}{CL\xspace}}
\ifthenelse{\boolean{cms@external}}{\providecommand{\NA}{\ensuremath{\cdots}}}{\providecommand{\NA}{\text{---}}}
\newcommand{\CLs}{\ensuremath{\mathrm{CL}_\mathrm{s}}\xspace}
\newcommand{\RTL}{\ensuremath{R_\mathrm{TL}}\xspace}
\cmsNoteHeader{SMP-13-015}
\title{Study of vector boson scattering and search for new physics in events with two same-sign leptons and two jets}

\date{\today}

\abstract{
A study of vector boson scattering in $\Pp\Pp$ collisions at
a center-of-mass energy of 8\TeV is presented. The data sample corresponds to an integrated luminosity of 19.4\fbinv collected with the CMS detector. Candidate events are selected
with exactly two leptons of the same charge, two jets
with large rapidity separation and high dijet mass, and moderate missing transverse energy. The signal region
is expected to be dominated by electroweak same-sign \PW-boson pair production.
The observation agrees with the standard model prediction.
The observed significance is 2.0 standard deviations,
where a significance of 3.1 standard deviations is expected based on the standard model. Cross section
measurements for $\PW^\pm\PW^\pm$ and
$\PW\cPZ$ processes in the fiducial region are reported. Bounds on the
structure of quartic vector-boson interactions are given in the
framework of dimension-eight effective field theory operators, as well as limits on the
production of doubly-charged Higgs bosons.
}

\hypersetup{%
pdfauthor={CMS Collaboration},%
pdftitle={Study of vector boson scattering and search for new physics in events with two same-sign leptons and two jets},%
pdfsubject={CMS},%
pdfkeywords={CMS, physics, VBS, VBF, doubly charged Higgs}}

\maketitle
Vector boson scattering (VBS) and quartic boson couplings are features
of the standard model (SM) that remain largely unexplored by the LHC
experiments.  The observation of a Higgs
boson~\cite{atlas-higgs,cms-higgs,CMSPaperCombinationLong}, in
accordance with a key prediction of the SM, motivates further study of
the mechanism of electroweak symmetry breaking through measurements of
VBS processes.  In the absence of the SM Higgs boson, the amplitudes
for these processes would increase as a function of center-of-mass
energy and ultimately violate
unitarity~\cite{PhysRevLett.38.883,PhysRevD.16.1519}.  The Higgs boson
actually observed by the LHC experiments may restore the unitarity,
although some scenarios of physics beyond the SM predict enhancements
for VBS through modifications to the Higgs sector or the presence of
additional resonances~\cite{PhysRevD.87.055017, PhysRevD.87.093005}.

\begin{figure}[htbp!]
  \begin{center}
   \includegraphics[width=0.23\textwidth]{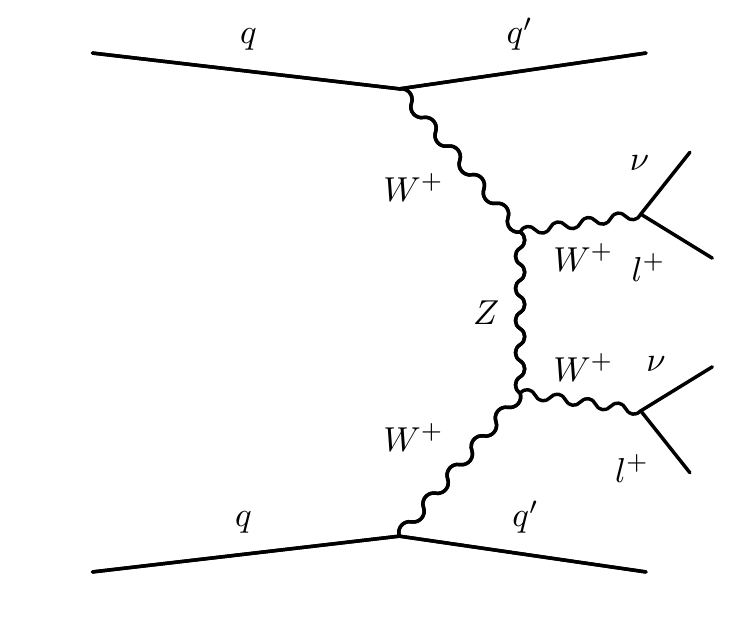}
   \includegraphics[width=0.23\textwidth]{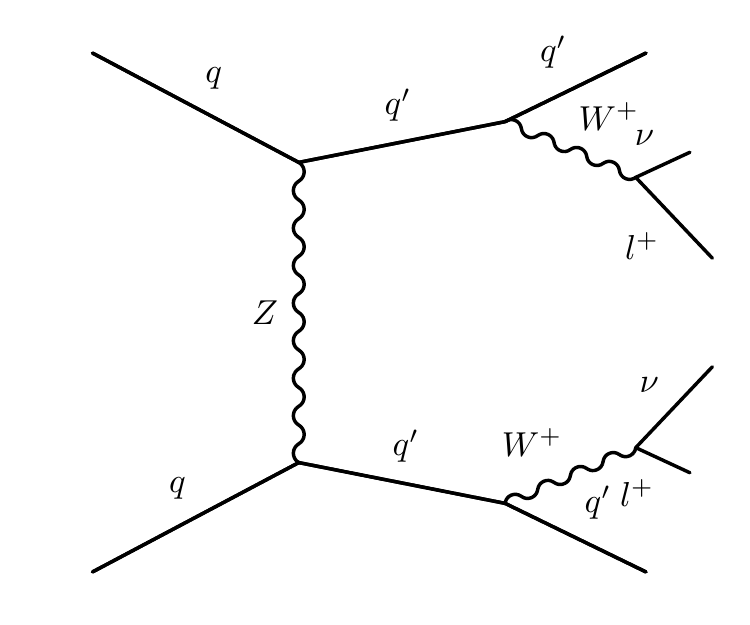}
   \includegraphics[width=0.23\textwidth]{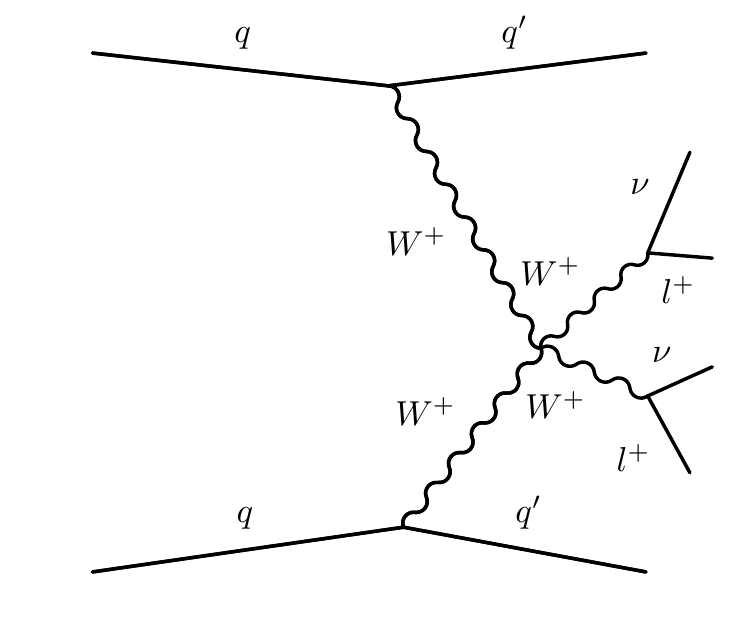}
   \includegraphics[width=0.23\textwidth]{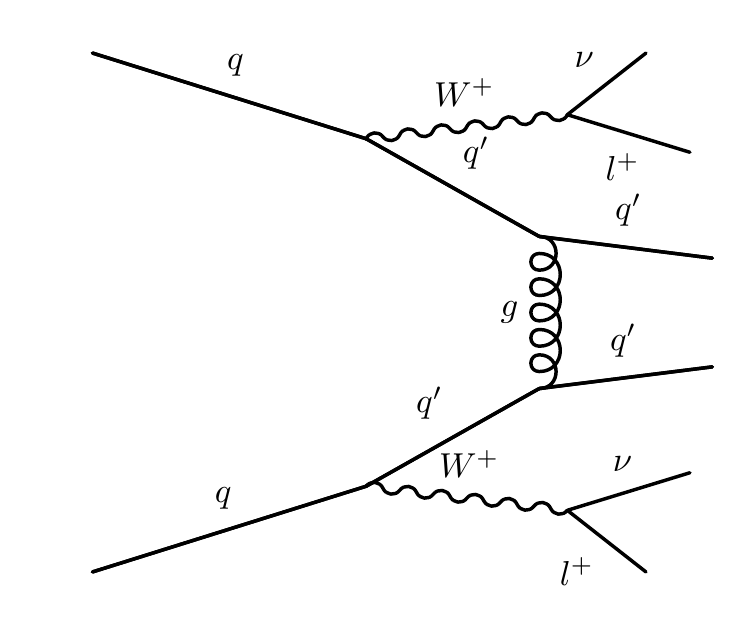}
    \caption{Representative Feynman diagrams for the electroweak and QCD induced same-sign $\PW$-boson pair production.}
    \label{fig:diagrams}
  \end{center}
\end{figure}

This letter presents a study of VBS in $\Pp\Pp$ collisions at $\sqrt{s}=8\TeV$.
The data sample corresponds to an
integrated luminosity of $19.4 \pm 0.5\fbinv$ collected with the CMS
detector~\cite{CMSdetector} at the LHC in 2012.
The aim of the analysis is find evidence for the electroweak production
of same-sign $\PW$-boson pair events. The strong production cross
section is reduced by the same-sign requirement, making the
experimental signature of same-sign dilepton events with two jets an
ideal topology for VBS studies.
Candidate events have exactly two identified leptons of the
same charge, two jets with large rapidity separation and dijet mass, and moderate
missing transverse energy. The final states considered are $\Pgmp
\Pgmp \Pgn_{\Pgm} \Pgn_{\Pgm} j j$, $\Pep \Pep \Pgn_{\Pe} \Pgn_{\Pe} j j$, $\Pep
\Pgmp \Pgn_{\Pe} \Pgn_{\Pgm} j j$, and their charge
conjugates and $\tau$-lepton decays to electrons
and muons.
Figure~\ref{fig:diagrams} shows representative Feynman diagrams for
the electroweak and QCD induced production.

The study of VBS presented here leads to
measurements of the production cross sections for $\PW^{\pm}\PW^{\pm}$
and $\PW\cPZ$ in a fiducial region. Evidence for electroweak
production has been reported by the ATLAS Collaboration~\cite{atlasww}.
Various extensions of the SM alter the couplings of vector bosons. An
excess of events could signal the presence of anomalous quartic gauge
couplings (AQGC)~\cite{aqgc_operators}. Doubly-charged Higgs bosons
are predicted in Higgs sectors beyond the SM where weak isotriplet
scalars are included~\cite{CE1, CE2}; they can be produced via
weak vector-boson fusion (VBF) and decay to pairs of same-sign $\PW$ bosons~\cite{1202.2014}.

The central feature of the CMS apparatus is a superconducting
solenoid, of 6\unit{m} internal diameter, providing a magnetic field of 3.8\unit{T}. Within
the field volume are a silicon pixel and strip tracker, a crystal
electromagnetic calorimeter, and a brass/scintillator hadron
calorimeter. Muons are measured in gas-ionization detectors embedded
in the steel flux-return yoke of the magnet.
The first level of the CMS trigger system, composed of custom hardware processors, is
designed to select the most interesting events within 3$\mus$, using
information from the calorimeters and muon detectors. The high level trigger processor farm further
reduces the event rate to a few hundred hertz before data storage.
Details of the CMS detector and its performance can be found elsewhere~\cite{CMSdetector}.

Several Monte Carlo (MC) event generators are used to simulate the signal and
background processes. The leading-order event
generator \MADGRAPH 5.2~\cite{Alwall:2014} is used to
produce event samples of diboson production via diagrams with two or fewer powers of $\alpha_s$ and up to six electroweak vertices. This includes two categories of diagrams: those with exactly two powers of $\alpha_s$ which we refer to as quantum chromodynamic (QCD) production and those with no powers of $\alpha_s$, which we refer to as electroweak (EW) production. The EW category includes diagrams with $\PW\PW\PW\PW$ quartic interactions and diagrams where two same-sign W bosons scatter through the exchange of a Higgs boson, a $\cPZ$ boson, or a photon. Double-parton scattering, triboson production, and doubly-charged Higgs boson production samples are also
generated using \MADGRAPH 5.2. Top-quark background processes are
generated with the next-to-leading-order event generator \POWHEG 1.0~\cite{POWHEG-V, POWHEG1, POWHEG2, POWHEG3}.
The set of parton distribution functions (PDFs) used is CTEQ6L~\cite{cteq66}
for \MADGRAPH and CT10~\cite{Lai:2010vv} for \POWHEG.
All event generators are interfaced
to \PYTHIA 6.4~\cite{pythia} for the showering of the partons and subsequent hadronization.
The \PYTHIA parameters for the underlying event were set according to the Z2* tune~\cite{z2star}.
The detector response is simulated by the \GEANTfour
package~\cite{Agostinelli:2002hh} using a detailed description of the CMS detector.
The average number of simultaneous proton-proton interactions per bunch
crossing in the 8\TeV data is approximately 21;
additional $\Pp\Pp$ interactions overlapping with the event of interest
are included in the simulated samples.
Collision events are selected by the trigger system requiring the presence
of one or two high transverse momentum (\pt) muons or electrons.
The trigger efficiency is greater than 99\% for events that pass all
other selection criteria explained below. A particle-flow algorithm~\cite{pf1,pf2} is used to reconstruct all
observable particles in the event. It combines all subdetector information
to reconstruct individual particles and identify them as charged
hadrons, neutral hadrons, photons and leptons.
The missing transverse energy $\MET$ is defined as
the magnitude of the negative vector sum of the transverse momenta of all reconstructed
particles (charged and neutral) in the event.

The selection of events aims to single out same-sign dilepton events with
the VBS topology while reducing the top quark,
Drell--Yan, and $\PW\cPZ$ background processes. All objects are selected
following the methods described in Ref.~\cite{HWW2013}.
To avoid bias, the number of events passing the selection was not evaluated until the analysis was complete.
Two same-sign lepton candidates, muons or electrons, with
$\pt > 20\GeV$ and $\abs{\eta} <2.4\,(2.5)$ for muons (electrons) are
required to be isolated from other reconstructed particles
 in a cone of $\Delta R = 0.3$, where $\Delta R = \sqrt{ \smash[b]{\Delta \eta^2 + \Delta \phi^2}}$.
Jets are reconstructed using the anti-\kt
clustering algorithm~\cite{antikt}  with a distance parameter $R=0.5$,
as implemented in the \textsc{fastjet}  package~\cite{Cacciari:fastjet1,Cacciari:fastjet2}.
Events are required to have at least two selected jets with $\ET>30\GeV$ and $\abs{\eta}<4.7$.
The VBS topology is targeted by requiring that the
two jets with leading $\pt$ have large dijet mass, $m_{jj} > 500\GeV$, and large pseudorapidity
separation, $\abs{\Delta \eta_{jj}} >
2.5$.

To suppress top-quark backgrounds ($\ttbar$ and $\cPqt\PW$), a top-quark veto
technique is used; it is based on the presence of a soft-muon in the event from the semileptonic
decay of the bottom-quark and on bottom-quark jet tagging criteria based on the impact
parameters of the constituent tracks~\cite{btag}. A minimum
dilepton mass,  $\mll > 50\GeV$, is required to reduce the $\PW\text{+jets}$ and top-quark background
processes.
To reduce the background from $\PW\cPZ$ production, events with a third, loosely identified
lepton with $\pt > 10\GeV$ are rejected. Drell--Yan events can be
selected if the charge of one lepton is measured incorrectly.
To reduce this background, $\abs{\mll - m_\cPZ } > 15\GeV$ is required for
$\Pe^{\pm} \Pe^{\pm}$ events.
The charge confusion in dimuon events is negligible. The  Drell--Yan background is further reduced by requiring $\MET > 40\GeV$.

The nonprompt lepton background originating from leptonic decays of heavy quarks, hadrons misidentified as leptons, and electrons from
photon conversions, is suppressed by the identification and isolation
requirements imposed on muons and electrons. The remaining
contribution from the nonprompt lepton background is estimated
directly from data. The efficiency for a predefined loose lepton-like
object to pass the full lepton selection, typically called the
``tight-to-loose ratio" (\RTL), is estimated in a control sample with
one additional lepton candidate that passes the standard lepton
selection criteria. To account for the dependence on kinematic
observables, this ratio is parameterized as a function of $\pt$  and
$\eta$. Systematic uncertainties are obtained by the application of \RTL to other control samples, accounting for the sample dependence in
the estimation of \RTL.
The $\PW\cPZ\to 3\ell\Pgn$ process is normalized in a data control
region by requiring
a third fully identified lepton with $\pt > 10\GeV$.
The contribution of opposite-sign dilepton events to the signal region is estimated
by applying data-to-simulation charge misidentification scale factors to simulated events
with two opposite-sign leptons. The charge-misidentification fraction
is estimated using $\cPZ$ boson events and is found to be between 0.1\%
and 0.3\% for electrons, while it is negligible for muons.

After the full selection, about 15\% of the background is due to the $\PW\cPZ\to 3\ell\Pgn$ process and about 75\% to nonprompt leptons.
Backgrounds from opposite-sign lepton pairs misreconstructed as same-sign
(``wrong-sign background"), $\PW\PW$ production via double
parton scattering (DPS), and triboson production ($\V\V\V$), which includes
top-pair plus boson processes, contribute less than 10\%.

\begin{figure}[bthp]
  \centering
 \includegraphics[width=0.45\textwidth]{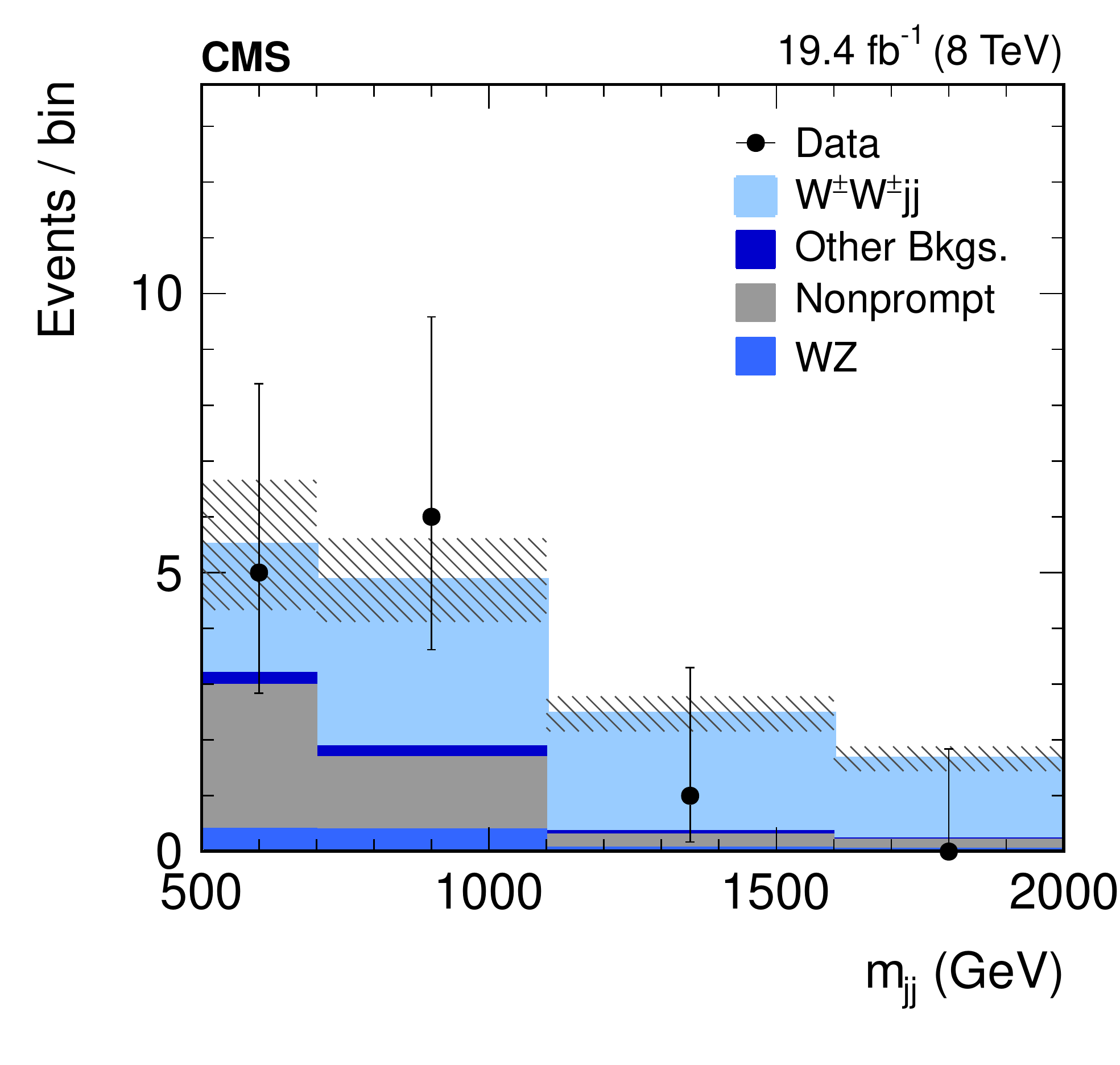}
 \includegraphics[width=0.45\textwidth]{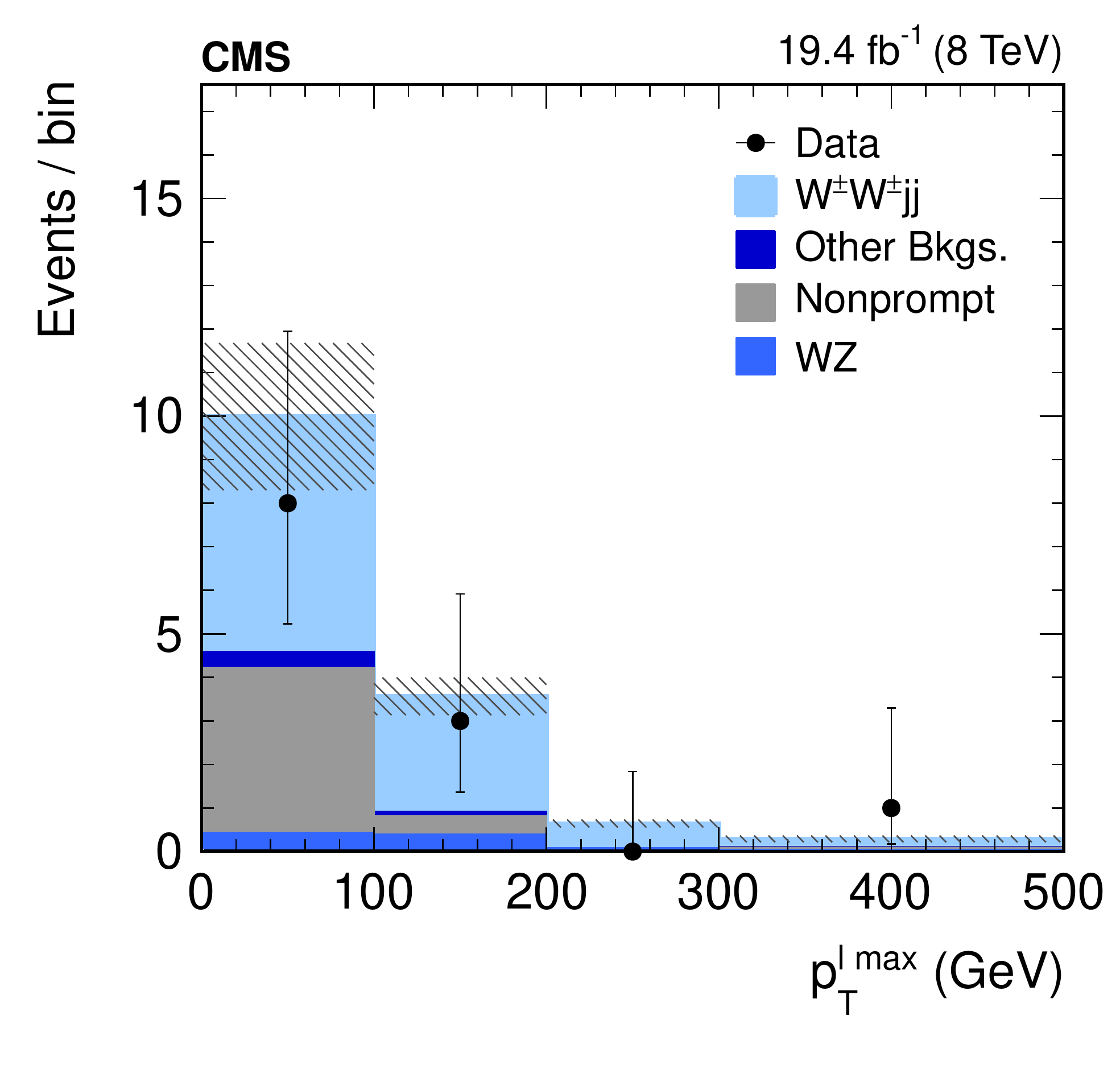}
    \caption{The distributions of $m_{jj}$ (\cmsLeft) and leading
      lepton $\pt$, $\ptlmax$,
      in the signal region (\cmsRight).
      The hatched bars include statistical and systematic
      uncertainties. The
       $\PWp\PWp$ and $\PWm\PWm$ candidates are combined
       in these distributions. The signal, $\PW^{\pm}\PW^{\pm} jj$, includes EW and
       QCD processes and their interference. The histograms for other backgrounds include the contributions
from wrong-sign events, DPS, and $\V\V\V$ processes.}
    \label{fig:signal_region}
\end{figure}

The expected signal and background yields are shown in
Table~\ref{tab:yields} for positive and negative pairs separately and
for their sum.
The signal corresponds to $\PW^{\pm}\PW^{\pm}$ production, including EW and QCD contributions,
and their interference, which amounts to approximately 10\%.
The EW processes constitute
85--90\% of the total signal contribution.
The $m_{jj}$ and leading-lepton $\pt$ distributions
for the signal and background processes are shown in
Fig.~\ref{fig:signal_region}.
In order to quantify the significance of the observation of the
production via VBS, a statistical analysis of the event yields
is performed in eight bins: four bins in $m_{jj}$ with two bins in the lepton charge.

\begin{table*}[bthp]
\centering
\topcaption{Signal and background yields after the full selection. Only statistical
uncertainties are reported. The signal, $\PW^{\pm}\PW^{\pm} jj$, includes EW and
QCD processes and their interference.}
\label{tab:yields}
\resizebox{\textwidth}{!}{
\begin{scotch}{ccccccccc}
                        & Nonprompt     & $\PW\cPZ$     & $\V\V\V$      & Wrong sign     & $\PW\PW$ DPS  & Total bkg.    & $\PW^{\pm}\PW^{\pm} jj$ & Data\\
  \hline
  $\PWp\PWp$      & 2.1 $\pm$ 0.6 & 0.6 $\pm$ 0.1 & 0.2 $\pm$ 0.1 & 0.1 $\pm$ 0.1  & 0.1 $\pm$ 0.1 & 3.1 $\pm$ 0.6 & 7.1 $\pm$ 0.1 & 10 \\
  $\PWm\PWm$      & 2.1 $\pm$ 0.5 & 0.4 $\pm$ 0.1 & 0.1 $\pm$ 0.1 &	\NA      &    \NA        & 2.6 $\pm$ 0.5 & 1.8 $\pm$ 0.1 & 2  \\
  \hline
  $\PW^{\pm}\PW^{\pm}$  & 4.2 $\pm$ 0.8 & 1.0 $\pm$ 0.1 & 0.3 $\pm$ 0.1 & 0.1 $\pm$ 0.1  & 0.1 $\pm$ 0.1 & 5.7 $\pm$ 0.8 & 8.9 $\pm$ 0.1 & 12 \\
\end{scotch}
}
\end{table*}

The signal efficiencies are estimated using simulated samples. In the
statistical analysis, shape and normalization uncertainties are considered.
The shape uncertainties are estimated by remaking the distribution of a given observable
after considering the systematic variations for each source of uncertainty.
The lepton trigger, reconstruction, and selection efficiencies are measured using $\dyll$~events that provide an
unbiased sample with high purity. The estimated uncertainty
is 2\% per lepton. The uncertainties due to the momentum scale for electrons
and muons are also taken into account and
contribute 2\%. The jet energy
scale and resolution uncertainties give rise to an uncertainty in the
yields of about 5\%. The uncertainty in the event selection efficiency
for events with neutrinos yielding genuine $\MET$ in the final state is
assessed and leads to an uncertainty of 2\%.
The uncertainty in the estimated event yields, which is related to the top-quark veto, is evaluated
by using a $\dyll$ sample with at least two reconstructed jets
and is found to be about 2\%. The statistical uncertainty in the yield of each bin and
for each process is also taken into account. The uncertainty of 2.6\% in the integrated
luminosity~\cite{lumiPAS2012} is considered for all simulated processes.
The normalization of the processes with misidentified leptons has a 36\% systematic
uncertainty~\cite{HWW2013}, which has two sources: the dependence on the sample
composition and the method used to estimate it. The $\PW\cPZ$ normalization
uncertainty is 35\%, dominated by the small number of events in the
trilepton control region. Theoretical uncertainties are estimated by varying
the renormalization and factorization scales up and down by a factor of two from
their nominal value in the event, and found to be 5\% for the signal
normalization and 50\% for the triboson background normalization.
A PDF uncertainty of 6--8\% in the normalization
of the signal and $\PW\cPZ$ processes is included.
The systematic uncertainties of the background
normalizations are taken into account using log-normal distributions.

The cross section is extracted for a fiducial signal region.
The fiducial region is defined
by requiring
two same-sign leptons with $\pt^{\ell}>10\GeV$ and $\abs{\eta_{\ell}}<2.5$,
two jets with $\pt^{j}>20\GeV$ and $\abs{\eta^{j}}<5.0$,
$m_{jj}>300\GeV$, and  $\abs{\Delta \eta_{jj}}>2.5$ and is less stringent than the event
selection for our signal region.
The measured cross section is corrected for the
acceptance in this region using the \MADGRAPH MC generator,
which is also used to estimate the theoretical cross section.
The acceptance ratio between the selected signal region and the fiducial
region is 36\% considering generator-level jet and lepton properties only. The overall
acceptance times efficiency is 7.9\%.

The \MADGRAPH prediction of the same-sign $\PW$-boson pair cross section
is corrected by a next-to-leading order to leading-order cross section ratio estimated using
\textsc{vbfnlo}~\cite{vbfnlo1,vbfnlo2,vbfnlo3}.
The fiducial cross section is found to be $\sigma_\text{fid}(\PW^{\pm}\PW^{\pm} jj) =
4.0^{+2.4}_{-2.0}\stat ^{+1.1}_{-1.0}\syst\unit{fb}$ with an
expectation of $5.8 \pm 1.2\unit{fb}$.

In addition to the dilepton same-sign signal region, a
$\PW\cPZ \to 3\ell\Pgn$ control region is studied by requiring
an additional lepton with $\pt$ larger than 10\GeV. This control
region allows the measurement of a fiducial cross section of the
$\PW\cPZ jj$ process and is $\sigma_\text{fid}(\PW\cPZ jj) = 10.8 \pm 4.0\stat \pm 1.3\syst\unit{fb}$ with an expectation of $14.4 \pm 4.0
\unit{fb}$. The fiducial region is defined in the same way
as for the $\PW\PW$ analysis, but requiring one more lepton with
$\pt^{\ell}>10\GeV$ and $\abs{\eta_{\ell}}<2.5$. The acceptance ratio
between the selected signal region and the fiducial region is 20\%
considering generator-level jet and lepton properties only. The
overall acceptance times efficiency is 3.6\%.

To compute the limits and significances, the \CLs~\cite{Read1,junkcls,LHC-HCG} construction is used.
The observed (expected) significance for the $\PW^{\pm}\PW^{\pm}jj$
process is 2.0$\,\sigma$ (3.1$\,\sigma$).
Considering the QCD component of the $\PW^{\pm}\PW^{\pm}jj$ events as background
and the EW component together with the EW-QCD interference as signal, the observed
(expected) signal significance reduces to 1.9$\,\sigma$ (2.9$\,\sigma$).

\begin{figure}[htbp]
\centering
    \includegraphics[width=0.45\textwidth]{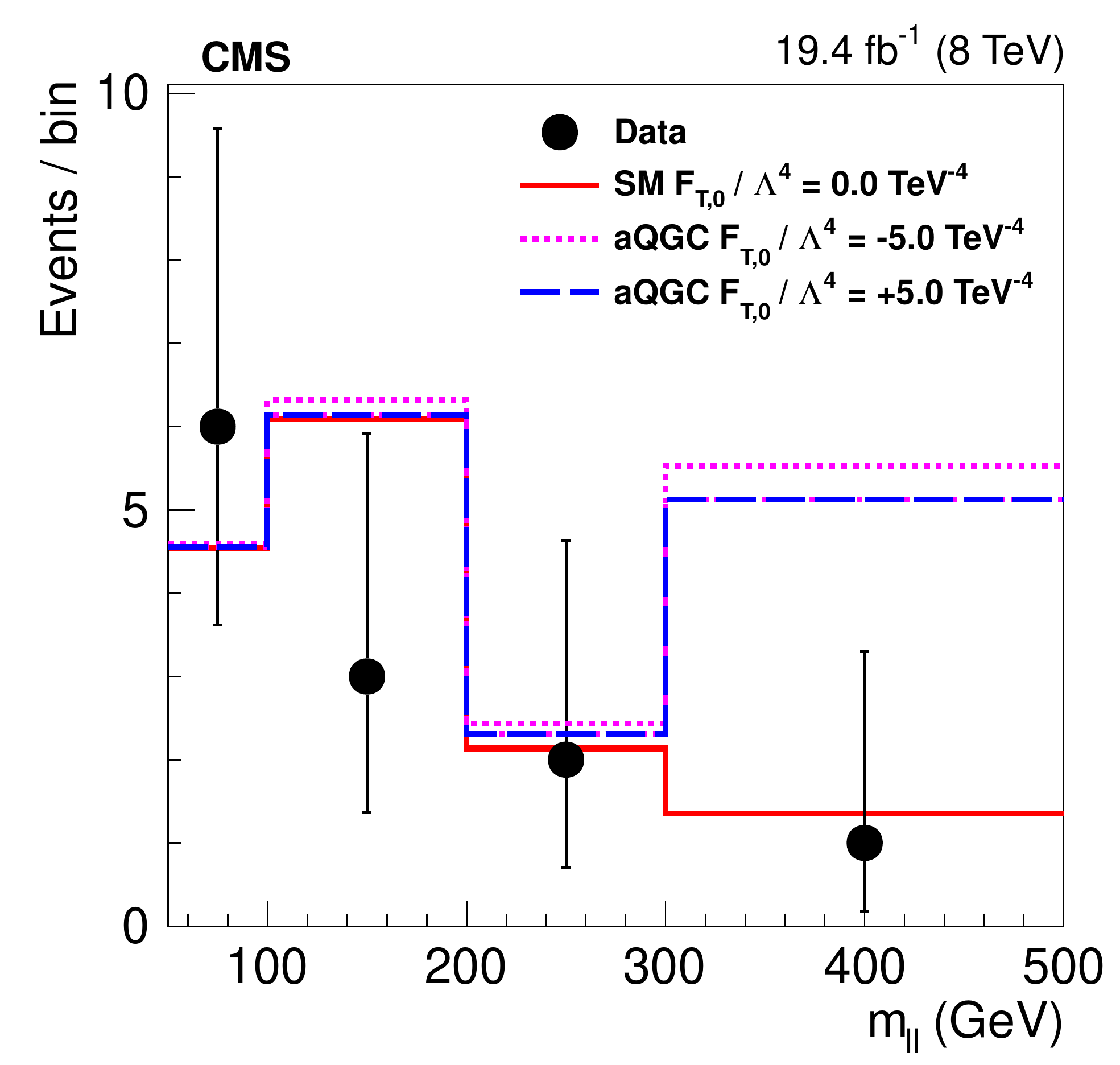}
    \includegraphics[width=0.45\textwidth]{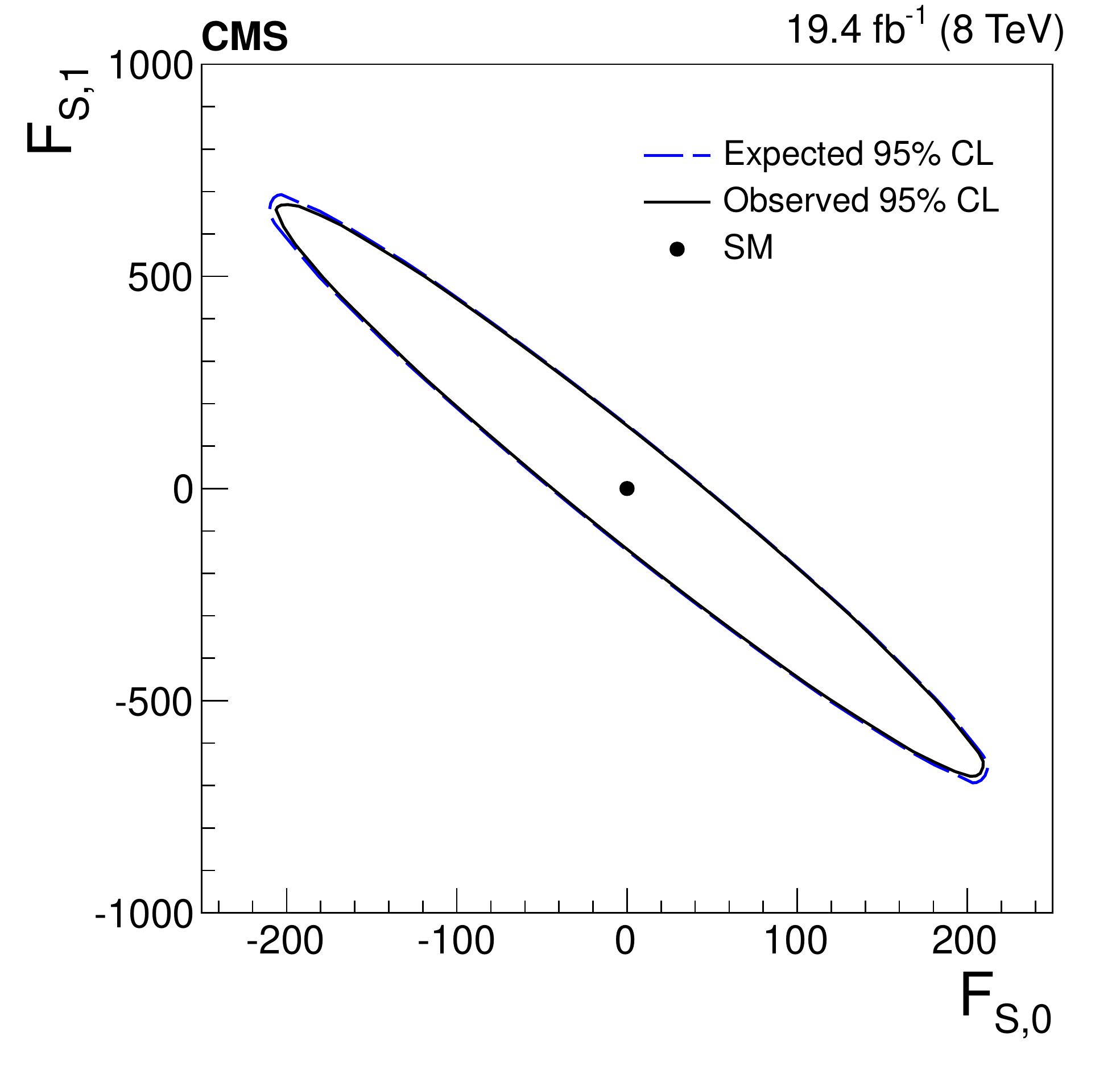}
    \caption{The $\mll$ distributions (\cmsLeft) after full selection with all
    SM backgrounds and $F_{T,0} / \Lambda^4= -5.0$, 0 (SM), and $5.0\TeV^{-4}$; the
    last bin includes overflow events. Observed and expected two-dimensional
    95\% \CL limits (\cmsRight) for $F_{S0}/ \Lambda^4$ and $F_{S1}/ \Lambda^4$.}
    \label{fig:aqgc_limit}
\end{figure}

Various extensions to the SM alter the couplings between vector bosons.
Ref. ~\cite{aqgc_operators} proposes nine independent C- and P-conserving dimension-eight
effective operators to modify the quartic couplings
between the weak gauge bosons. The variable $\mll$ is more sensitive
to AQGCs than $\ptlmax$, $m_{\ell\ell jj}$, and $m_{jj}$.
Fig.~\ref{fig:aqgc_limit} (\cmsLeft) shows the expected $\mll$
distribution for three values of $F_{T,0} / \Lambda^4$;
$\Lambda$ is the scale of new physics and $F_{T,0}$ is the coefficient of one of the nine
effective operators.
The observed and expected upper and lower limits at 95\% confidence
level (\CL) on the nine coefficients are shown
in Table~\ref{tab:aqgc_limits}, where all the results are obtained by varying
the effective operators one by one. The effect of possible AQGCs on
the $\PW\cPZ$ process in the signal region is negligible. Some operators for anomalous quartic gauge boson couplings may lead to tree-level unitarity violation. We also report the values
of the operator coefficient for which unitarity is restored at the scale
of 8\TeV, the unitarity limit.
In addition to the limits on individual operator coefficients, the expected and
observed two-dimensional 95\% \CL limits on $F_{S,0} / \Lambda^4$ and $F_{S,1} / \Lambda^4$
are presented in Fig.~\ref{fig:aqgc_limit} (\cmsRight): a linear
combination of those operators leads to a scaling of the SM cross section.

\begin{table*}[htbp]
\topcaption{Observed and expected upper and lower limits at 95\% \CL limits on the nine dimension-eight
  operators that effect quartic couplings between the weak gauge
  bosons. Limits from unitarity are reported. The units are $\TeV^{-4}$.}
\label{tab:aqgc_limits}
\centering
\begin{scotch}{cccccc}
 Operator coefficient & Exp. lower  & Exp. upper  &  Obs. lower &
 Obs. upper & Unitarity limit \\
\hline
$F_{S,0} / \Lambda^4$ & $-42$ & $43$ & $-38$ & $40$ & $0.016$\\
$F_{S,1} / \Lambda^4 $ & $-129$ & $131$ & $-118$ & $120$ & $0.050$\\
$F_{M,0} / \Lambda^4 $ & $-35$ & $35$ & $-33$ & $32$ & $80$\\
$F_{M,1} / \Lambda^4$ & $-49$ & $51$ & $-44$ & $47$ & $205$\\
$F_{M,6} / \Lambda^4$ & $-70$ & $69$ & $-65$ & $63$  & $160$\\
$F_{M,7} / \Lambda^4$ & $-76$ & $73$ & $-70$ & $66$  & $105$\\
$F_{T,0} / \Lambda^4$ & $-4.6$ & $4.9$ & $-4.2$ & $4.6$ & $0.027$\\
$F_{T,1} / \Lambda^4$ & $-2.1$ & $2.4$ & $-1.9$ & $2.2$ & $0.022$\\
$F_{T,2} / \Lambda^4$ & $-5.9$ & $7.0$ & $-5.2$ & $6.4$ & $0.08$\\
\end{scotch}
\end{table*}

\begin{figure}[htbp]
\centering
    \includegraphics[width=0.45\textwidth]{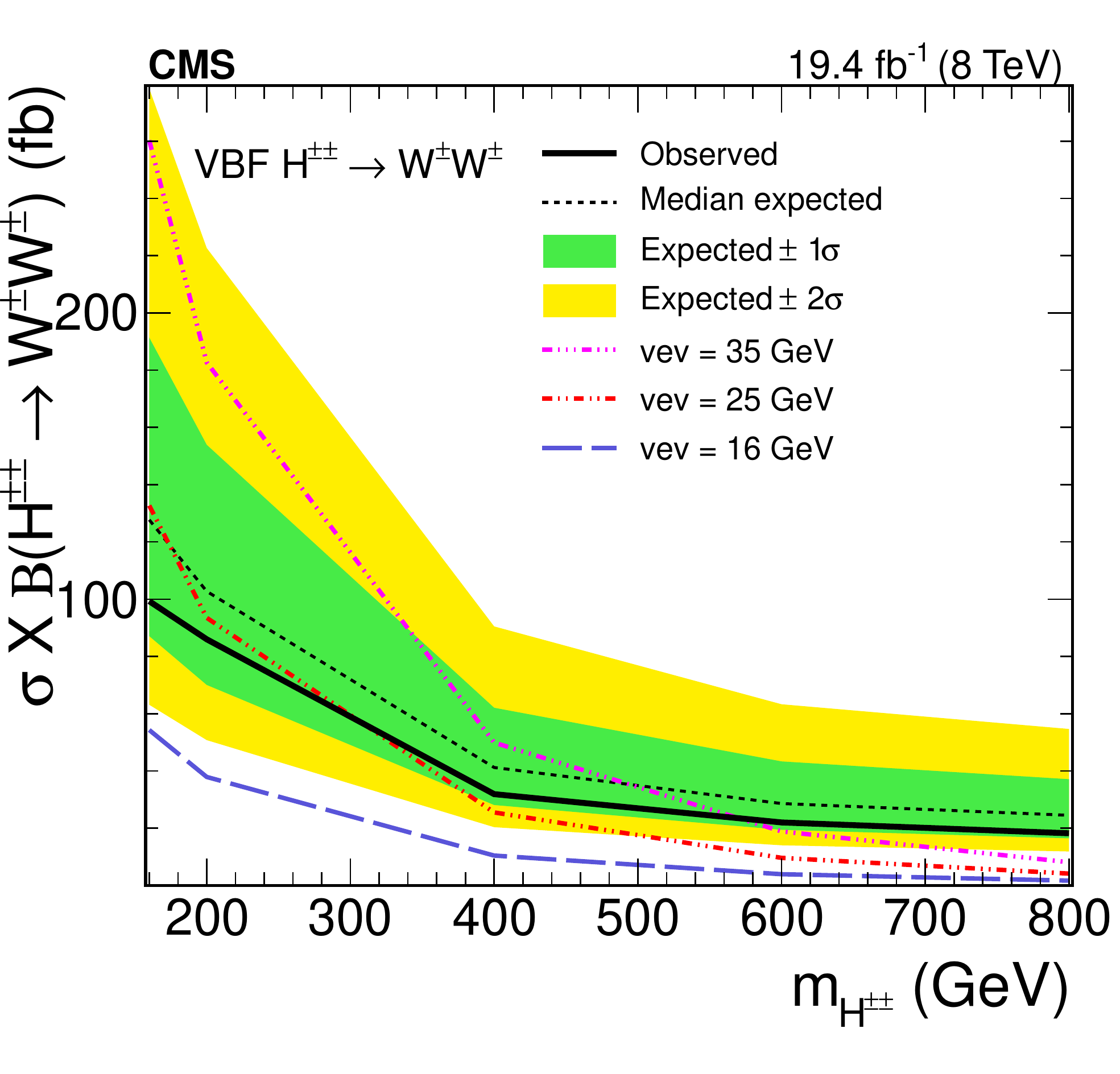}
    \caption{Expected and observed 95\% \CL upper
       limits on the cross section times branching fraction,
        $\sigma_{\PH^{\pm\pm}}\mathcal{B}(\PH^{\pm\pm} \to
       \PW^{\pm}\PW^{\pm})$. Theoretical cross sections for three
       values of the vacuum expectation value (vev) are overlaid.}
    \label{fig:ana_hpp}
\end{figure}

Doubly-charged Higgs bosons are predicted in models that contains a
Higgs triplet field. Some of these scenarios predict same-sign dilepton events from $\PW^\pm\PW^\pm$
decays with a VBF topology.
The  cross section for VBF production of $\PH^{\pm\pm}$ and decay to
$\PW^\pm\PW^\pm$ is directly proportional to the vacuum expectation value of the triplet.
The remaining five parameters in the
model of the Higgs potential are adjusted to get the given $m_{\PH^{\pm\pm}}$ hypothesis
while requiring one of the scalar singlets to have a
mass of 125\GeV. The Georgi--Machacek model of Higgs
triplets~\cite{GMH} is considered. For $m_{\PH^{\pm\pm}} = 200\,(800)\GeV$ the
following parameters are used: $\lambda_1 = 1$, $\lambda_2 = 1$, $\lambda_3 = 1$,
$\lambda_4 = 2.37\,(4)$, and $\lambda_5 = 0.432\,(7.26)$.
By using the $m_{jj}$ distribution, 95\% \CL upper limits on $\sigma_{\PH^{\pm\pm}}
\mathcal{B}(\PH^{\pm\pm} \to \PW^{\pm}\PW^{\pm})$ are derived as shown in
Fig.~\ref{fig:ana_hpp}. The experimental results are overlaid with theoretical cross sections for three
       values of the vacuum expectation value.

In summary, a study of vector boson scattering in $\Pp\Pp$ collisions at
$\sqrt{s}=8\TeV$ has been presented based on a data sample corresponding to an
integrated luminosity of 19.4\fbinv.
Candidate events are selected with exactly two leptons of the same charge, two jets
with large rapidity separation and dijet mass, and moderate missing
transverse energy. The signal region is expected to be dominated by
electroweak same-sign $\PW$-boson pair production.  The
observation agrees with the standard model prediction.  The observed
significance is 2.0 standard deviations, where a significance of 3.1
standard deviations is expected based on the standard model. Cross
section measurements for $\PW^{\pm}\PW^{\pm}$ and
$\PW\cPZ$ processes in the fiducial region are reported. Bounds on the
structure of quartic vector-boson interactions are given in the
framework of dimension-eight effective field theory operators, as well as limits on the
production of doubly-charged Higgs bosons.
\begin{acknowledgments}
We congratulate our colleagues in the CERN accelerator departments for the excellent performance of the LHC and thank the technical and administrative staffs at CERN and at other CMS institutes for their contributions to the success of the CMS effort. In addition, we gratefully acknowledge the computing centers and personnel of the Worldwide LHC Computing Grid for delivering so effectively the computing infrastructure essential to our analyses. Finally, we acknowledge the enduring support for the construction and operation of the LHC and the CMS detector provided by the following funding agencies: BMWFW and FWF (Austria); FNRS and FWO (Belgium); CNPq, CAPES, FAPERJ, and FAPESP (Brazil); MES (Bulgaria); CERN; CAS, MoST, and NSFC (China); COLCIENCIAS (Colombia); MSES and CSF (Croatia); RPF (Cyprus); MoER, ERC IUT and ERDF (Estonia); Academy of Finland, MEC, and HIP (Finland); CEA and CNRS/IN2P3 (France); BMBF, DFG, and HGF (Germany); GSRT (Greece); OTKA and NIH (Hungary); DAE and DST (India); IPM (Iran); SFI (Ireland); INFN (Italy); NRF and WCU (Republic of Korea); LAS (Lithuania); MOE and UM (Malaysia); CINVESTAV, CONACYT, SEP, and UASLP-FAI (Mexico); MBIE (New Zealand); PAEC (Pakistan); MSHE and NSC (Poland); FCT (Portugal); JINR (Dubna); MON, RosAtom, RAS and RFBR (Russia); MESTD (Serbia); SEIDI and CPAN (Spain); Swiss Funding Agencies (Switzerland); MST (Taipei); ThEPCenter, IPST, STAR and NSTDA (Thailand); TUBITAK and TAEK (Turkey); NASU and SFFR (Ukraine); STFC (United Kingdom); DOE and NSF (USA).
\end{acknowledgments}
\bibliography{auto_generated}   

\cleardoublepage \appendix\section{The CMS Collaboration \label{app:collab}}\begin{sloppypar}\hyphenpenalty=5000\widowpenalty=500\clubpenalty=5000\input{SMP-13-015-authorlist.tex}\end{sloppypar}
\end{document}

%% file: SMP-13-015-authorlist.tex
\textbf{Yerevan Physics Institute,  Yerevan,  Armenia}\\*[0pt]
V.~Khachatryan, A.M.~Sirunyan, A.~Tumasyan
\vskip\cmsinstskip
\textbf{Institut f\"{u}r Hochenergiephysik der OeAW,  Wien,  Austria}\\*[0pt]
W.~Adam, T.~Bergauer, M.~Dragicevic, J.~Er\"{o}, M.~Friedl, R.~Fr\"{u}hwirth\cmsAuthorMark{1}, V.M.~Ghete, C.~Hartl, N.~H\"{o}rmann, J.~Hrubec, M.~Jeitler\cmsAuthorMark{1}, W.~Kiesenhofer, V.~Kn\"{u}nz, M.~Krammer\cmsAuthorMark{1}, I.~Kr\"{a}tschmer, D.~Liko, I.~Mikulec, D.~Rabady\cmsAuthorMark{2}, B.~Rahbaran, H.~Rohringer, R.~Sch\"{o}fbeck, J.~Strauss, W.~Treberer-Treberspurg, W.~Waltenberger, C.-E.~Wulz\cmsAuthorMark{1}
% ...
\vskip\cmsinstskip
% ...
1:~~Also at Vienna University of Technology, Vienna, Austria\\
2:~~Also at CERN, European Organization for Nuclear Research, Geneva, Switzerland\\
% ...


%% file: SMP-13-015_temp.bbl
\providecommand{\href}[2]{#2}\begingroup\raggedright\begin{thebibliography}{10}%
\makeatletter
\providecommand{\hrefCMSnoop }[0]{\@secondoftwo}%
\makeatother
\providecommand{\doi}{\texttt{doi:}\begingroup \urlstyle{tt}\Url}

\bibitem{atlas-higgs}
\hrefCMSnoop {}{{ATLAS} Collaboration, ``{Observation of a new particle in the
  search for the Standard Model Higgs boson with the ATLAS detector at the
  LHC}'',} \textit{ Phys. Lett. B} \textbf{ 716} (2012) 1,
  \href{http://dx.doi.org/10.1016/j.physletb.2012.08.020}{\doi{10.1016/j.physletb.2012.08.020}},
  \href{http://www.arXiv.org/abs/1207.7214}{\texttt{ arXiv:1207.7214}}.

\bibitem{cms-higgs}
\hrefCMSnoop {}{{CMS} Collaboration, ``{Observation of a new boson at a mass of
  125 GeV with the CMS experiment at the LHC}'',} \textit{ Phys. Lett. B}
  \textbf{ 716} (2012) 30,
  \href{http://dx.doi.org/10.1016/j.physletb.2012.08.021}{\doi{10.1016/j.physletb.2012.08.021}},
  \href{http://www.arXiv.org/abs/1207.7235}{\texttt{ arXiv:1207.7235}}.

\bibitem{CMSPaperCombinationLong}
\hrefCMSnoop {}{{CMS} Collaboration, ``{Observation of a new boson with mass
  near 125 GeV in pp collisions at $\sqrt{s} = 7$ and 8 TeV}'',} \textit{ JHEP}
  \textbf{ 06} (2013) 081,
  \href{http://dx.doi.org/10.1007/JHEP06(2013)081}{\doi{10.1007/JHEP06(2013)081}},
  \href{http://www.arXiv.org/abs/1303.4571}{\texttt{ arXiv:1303.4571}}.

\bibitem{PhysRevLett.38.883}
\hrefCMSnoop {}{B.~W. Lee, C.~Quigg, and H.~B. Thacker, ``Strength of Weak
  Interactions at Very High Energies and the Higgs Boson Mass'',} \textit{
  Phys. Rev. Lett.} \textbf{ 38} (1977) 883,
  \href{http://dx.doi.org/10.1103/PhysRevLett.38.883}{\doi{10.1103/PhysRevLett.38.883}}.

\bibitem{PhysRevD.16.1519}
\hrefCMSnoop {}{B.~W. Lee, C.~Quigg, and H.~B. Thacker, ``Weak interactions at
  very high energies: The role of the Higgs-boson mass'',} \textit{ Phys. Rev.
  D} \textbf{ 16} (1977) 1519,
  \href{http://dx.doi.org/10.1103/PhysRevD.16.1519}{\doi{10.1103/PhysRevD.16.1519}}.

\bibitem{PhysRevD.87.055017}
\hrefCMSnoop {}{D.~Espriu and B.~Yencho, ``Longitudinal $WW$ scattering in
  light of the ``Higgs boson'' discovery'',} \textit{ Phys. Rev. D} \textbf{
  87} (2013) 055017,
  \href{http://dx.doi.org/10.1103/PhysRevD.87.055017}{\doi{10.1103/PhysRevD.87.055017}},
  \href{http://www.arXiv.org/abs/1212.4158}{\texttt{ arXiv:1212.4158}}.

\bibitem{PhysRevD.87.093005}
\hrefCMSnoop {}{J.~Chang, K.~Cheung, C.-T. Lu, and T.-C. Yuan, ``{$WW$
  scattering in the era of post-Higgs-boson discovery}'',} \textit{ Phys. Rev.
  D} \textbf{ 87} (2013) 093005,
  \href{http://dx.doi.org/10.1103/PhysRevD.87.093005}{\doi{10.1103/PhysRevD.87.093005}},
  \href{http://www.arXiv.org/abs/1303.6335}{\texttt{ arXiv:1303.6335}}.

\bibitem{CMSdetector}
\hrefCMSnoop {}{{CMS} Collaboration, ``The {CMS} experiment at the {CERN
  LHC}'',} \textit{ JINST} \textbf{ 3} (2008) S08004,
  \href{http://dx.doi.org/10.1088/1748-0221/3/08/S08004}{\doi{10.1088/1748-0221/3/08/S08004}}.

\bibitem{atlasww}
\hrefCMSnoop {}{{ATLAS} Collaboration, ``{Evidence for Electroweak Production
  of $W^{\pm}W^{\pm}jj$ in $pp$ Collisions at $\sqrt{s}=8$ TeV with the ATLAS
  Detector}'',} \textit{ Phys. Rev. Lett.} \textbf{ 113} (2014) 141803,
  \href{http://dx.doi.org/10.1103/PhysRevLett.113.141803}{\doi{10.1103/PhysRevLett.113.141803}},
\href{http://www.arXiv.org/abs/1405.6241}{\texttt{ arXiv:1405.6241}}.

\bibitem{aqgc_operators}
\hrefCMSnoop {}{{O. J. P. {\'E}boli, M. C. Gonzalez-Garcia, and J. K.
  Mizukoshi}, ``{p p $\to j j e^{\pm}\mu^{\pm} \nu \nu$ and $j j
  e^{\mp}\mu^{\pm} \nu \nu$ at O($\alpha^6_{em}$) and O($\alpha^4_{em}
  \alpha^2_s$) for the study of the quartic electroweak gauge boson vertex at
  CERN LHC}'',} \textit{ Phys. Rev. D} \textbf{ 74} (2006) 073005,
  \href{http://dx.doi.org/10.1103/PhysRevD.74.073005}{\doi{10.1103/PhysRevD.74.073005}},
  \href{http://www.arXiv.org/abs/hep-ph/0606118}{\texttt{
  arXiv:hep-ph/0606118}}.

\bibitem{CE1}
\hrefCMSnoop {}{{Ch. Englert, E. Re, and M. Spannowsky}, ``Triplet Higgs boson
  collider phenomenology after the LHC'',} \textit{ Phys. Rev. D} \textbf{ 87}
  (2013) 095014,
  \href{http://dx.doi.org/10.1103/PhysRevD.87.095014}{\doi{10.1103/PhysRevD.87.095014}},
  \href{http://www.arXiv.org/abs/1302.6505}{\texttt{ arXiv:1302.6505}}.

\bibitem{CE2}
\hrefCMSnoop {}{{Ch. Englert, E. Re, and M. Spannowsky}, ``Pinning down Higgs
  triplets at the LHC'',} \textit{ Phys. Rev. D} \textbf{ 88} (2013) 035024,
  \href{http://dx.doi.org/10.1103/PhysRevD.88.035024}{\doi{10.1103/PhysRevD.88.035024}},
  \href{http://www.arXiv.org/abs/1306.6228}{\texttt{ arXiv:1306.6228}}.

\bibitem{1202.2014}
\hrefCMSnoop {}{{Ch.-W. Chiang, T. Nomura, and K. Tsumura}, ``Search for doubly
  charged Higgs bosons using the same-sign diboson mode at the LHC'',} \textit{
  Phys. Rev. D} \textbf{ 85} (2012) 095023,
  \href{http://dx.doi.org/10.1103/PhysRevD.85.095023}{\doi{10.1103/PhysRevD.85.095023}},
  \href{http://www.arXiv.org/abs/1202.2014}{\texttt{ arXiv:1202.2014}}.

\bibitem{Alwall:2014}
\hrefCMSnoop {}{{J. Alwall et al.}, ``The automated computation of tree-level
  and next-to-leading order differential cross sections, and their matching to
  parton shower simulations'',} \textit{ JHEP} \textbf{ 07} (2014) 079,
  \href{http://dx.doi.org/10.1007/JHEP07(2014)079}{\doi{10.1007/JHEP07(2014)079}},
\href{http://www.arXiv.org/abs/1405.0301}{\texttt{ arXiv:1405.0301}}.

\bibitem{POWHEG-V}
\hrefCMSnoop {}{S.~Alioli, P.~Nason, C.~Oleari, and E.~Re, ``NLO vector-boson
  production matched with shower in POWHEG'',} \textit{ JHEP} \textbf{ 07}
  (2008) 060,
  \href{http://dx.doi.org/10.1088/1126-6708/2008/07/060}{\doi{10.1088/1126-6708/2008/07/060}},
  \href{http://www.arXiv.org/abs/0805.4802}{\texttt{ arXiv:0805.4802}}.

\bibitem{POWHEG1}
\hrefCMSnoop {}{P.~Nason, ``A new method for combining NLO QCD with shower
  Monte Carlo algorithms'',} \textit{ JHEP} \textbf{ 11} (2004) 040,
  \href{http://dx.doi.org/10.1088/1126-6708/2004/11/040}{\doi{10.1088/1126-6708/2004/11/040}},
  \href{http://www.arXiv.org/abs/hep-ph/0409146}{\texttt{
  arXiv:hep-ph/0409146}}.

\bibitem{POWHEG2}
\hrefCMSnoop {}{{S. Frixione, P. Nason, and C. Oleari}, ``Matching {NLO QCD}
  computations with parton shower simulations: the {POWHEG} method'',} \textit{
  JHEP} \textbf{ 11} (2007) 070,
  \href{http://dx.doi.org/10.1088/1126-6708/2007/11/070}{\doi{10.1088/1126-6708/2007/11/070}},
  \href{http://www.arXiv.org/abs/0709.2092}{\texttt{ arXiv:0709.2092}}.

\bibitem{POWHEG3}
\hrefCMSnoop {}{S.~Alioli, P.~Nason, C.~Oleari, and E.~Re, ``{A general
  framework for implementing NLO calculations in shower Monte Carlo programs:
  the POWHEG BOX}'',} \textit{ JHEP} \textbf{ 06} (2010) 043,
  \href{http://dx.doi.org/10.1007/JHEP06(2010)043}{\doi{10.1007/JHEP06(2010)043}},
\href{http://www.arXiv.org/abs/1002.2581}{\texttt{ arXiv:1002.2581}}.

\bibitem{cteq66}
H.-L. Lai\hrefCMSnoop {}{ {et~al.}, ``Uncertainty induced by {QCD} coupling in
  the {CTEQ} global analysis of parton distributions'',} \textit{ Phys. Rev. D}
  \textbf{ 82} (2010) 054021,
  \href{http://dx.doi.org/10.1103/PhysRevD.82.054021}{\doi{10.1103/PhysRevD.82.054021}},
  \href{http://www.arXiv.org/abs/1004.4624}{\texttt{ arXiv:1004.4624}}.

\bibitem{Lai:2010vv}
H.-L. Lai\hrefCMSnoop {}{ {et~al.}, ``{New parton distributions for collider
  physics}'',} \textit{ Phys. Rev. D} \textbf{ 82} (2010) 074024,
  \href{http://dx.doi.org/10.1103/PhysRevD.82.074024}{\doi{10.1103/PhysRevD.82.074024}},
  \href{http://www.arXiv.org/abs/1007.2241}{\texttt{ arXiv:1007.2241}}.

\bibitem{pythia}
\hrefCMSnoop {}{T.~Sj{\"o}strand, S.~Mrenna, and P.~Skands, ``PYTHIA 6.4
  physics and manual'',} \textit{ JHEP} \textbf{ 05} (2006) 026,
  \href{http://dx.doi.org/10.1088/1126-6708/2006/05/026}{\doi{10.1088/1126-6708/2006/05/026}},
  \href{http://www.arXiv.org/abs/hep-ph/0603175}{\texttt{
  arXiv:hep-ph/0603175}}.

\bibitem{z2star}
\hrefCMSnoop {}{{CMS} Collaboration, ``{Measurement of the Underlying Event
  Activity at the LHC with $\sqrt{s}= 7$ TeV and Comparison with $\sqrt{s} =
  0.9$ TeV}'',} \textit{ JHEP} \textbf{ 09} (2011) 109,
  \href{http://dx.doi.org/10.1007/JHEP09(2011)109}{\doi{10.1007/JHEP09(2011)109}},
  \href{http://www.arXiv.org/abs/1107.0330}{\texttt{ arXiv:1107.0330}}.

\bibitem{Agostinelli:2002hh}
\hrefCMSnoop {}{{GEANT4} Collaboration, ``{GEANT4}---a simulation toolkit'',}
  \textit{ Nucl. Instrum. Meth. A} \textbf{ 506} (2003) 250,
  \href{http://dx.doi.org/10.1016/S0168-9002(03)01368-8}{\doi{10.1016/S0168-9002(03)01368-8}}.

\bibitem{pf1}
\href {http://cdsweb.cern.ch/record/1194487}{{CMS} Collaboration,
  ``Particle--Flow Event Reconstruction in {CMS} and Performance for Jets,
  Taus, and {\MET}'',} CMS Physics Analysis Summary CMS-PAS-PFT-09-001, 2009.

\bibitem{pf2}
\href {http://cdsweb.cern.ch/record/1247373}{{CMS} Collaboration,
  ``Commissioning of the Particle-flow Event Reconstruction with the first LHC
  collisions recorded in the CMS detector'',} CMS Physics Analysis Summary
  CMS-PAS-PFT-10-002, 2010.

\bibitem{HWW2013}
\hrefCMSnoop {}{{CMS} Collaboration, ``Measurement of Higgs boson production
  and properties in the WW decay channel with leptonic final states'',}
  \textit{ JHEP} \textbf{ 01} (2014) 096,
  \href{http://dx.doi.org/10.1007/JHEP01(2014)096}{\doi{10.1007/JHEP01(2014)096}},
  \href{http://www.arXiv.org/abs/1312.1129}{\texttt{ arXiv:1312.1129}}.

\bibitem{antikt}
\hrefCMSnoop {}{M.~Cacciari, G.~P. Salam, and G.~Soyez, ``{The anti-${k_t}$ jet
  clustering algorithm}'',} \textit{ JHEP} \textbf{ 04} (2008) 063,
  \href{http://dx.doi.org/10.1088/1126-6708/2008/04/063}{\doi{10.1088/1126-6708/2008/04/063}},
\href{http://www.arXiv.org/abs/0802.1189}{\texttt{ arXiv:0802.1189}}.

\bibitem{Cacciari:fastjet1}
\hrefCMSnoop {}{M.~Cacciari, G.~P. Salam, and G.~Soyez, ``FastJet user
  manual'',} \textit{ Eur. Phys. J. C} \textbf{ 72} (2012) 1896,
  \href{http://dx.doi.org/10.1140/epjc/s10052-012-1896-2}{\doi{10.1140/epjc/s10052-012-1896-2}},
  \href{http://www.arXiv.org/abs/1111.6097}{\texttt{ arXiv:1111.6097}}.

\bibitem{Cacciari:fastjet2}
\hrefCMSnoop {}{M.~Cacciari and G.~P. Salam, ``{Dispelling the $N^{3}$ myth for
  the ${k_t}$ jet-finder}'',} \textit{ Phys. Lett. B} \textbf{ 641} (2006) 57,
  \href{http://dx.doi.org/10.1016/j.physletb.2006.08.037}{\doi{10.1016/j.physletb.2006.08.037}},
  \href{http://www.arXiv.org/abs/hep-ph/0512210}{\texttt{
  arXiv:hep-ph/0512210}}.

\bibitem{btag}
\hrefCMSnoop {}{{CMS} Collaboration, ``Identification of b-quark jets with the
  {CMS} experiment'',} \textit{ JINST} \textbf{ 8} (2012) P04013,
  \href{http://dx.doi.org/10.1088/1748-0221/8/04/P04013}{\doi{10.1088/1748-0221/8/04/P04013}},
  \href{http://www.arXiv.org/abs/1211.4462}{\texttt{ arXiv:1211.4462}}.

\bibitem{lumiPAS2012}
\href {http://cdsweb.cern.ch/record/1598864}{{CMS} Collaboration, ``CMS
  Luminosity Based on Pixel Cluster Counting - Summer 2013 Update'',} CMS
  Physics Analysis Summary CMS-PAS-LUM-13-001, 2013.

\bibitem{vbfnlo1}
J.~Baglio\hrefCMSnoop {}{ {et~al.}, ``Release Note - VBFNLO 2.7.0'',} (2014).
  \href{http://www.arXiv.org/abs/1404.3940}{\texttt{ arXiv:1404.3940}}.

\bibitem{vbfnlo2}
\hrefCMSnoop {}{J.~Baglio {et~al.}, ``VBFNLO: A parton level Monte Carlo for
  processes with electroweak bosons -- Manual for Version 2.7.0'',} (2011).
  \href{http://www.arXiv.org/abs/1107.4038}{\texttt{ arXiv:1107.4038}}.

\bibitem{vbfnlo3}
\hrefCMSnoop {}{K.~Arnold {et~al.}, ``VBFNLO: A parton level Monte Carlo for
  processes with electroweak bosons'',} \textit{ Comput. Phys. Commun.}
  \textbf{ 180} (2009) 1661,
  \href{http://dx.doi.org/10.1016/j.cpc.2009.03.006}{\doi{10.1016/j.cpc.2009.03.006}},
  \href{http://www.arXiv.org/abs/0811.4559}{\texttt{ arXiv:0811.4559}}.

\bibitem{Read1}
\hrefCMSnoop {}{A.~L. Read, ``Presentation of search results: the $CL_s$
  technique'',} \textit{ J. Phys. G} \textbf{ 28} (2002) 2693,
  \href{http://dx.doi.org/10.1088/0954-3899/28/10/313}{\doi{10.1088/0954-3899/28/10/313}}.

\bibitem{junkcls}
\hrefCMSnoop {}{T.~Junk, ``{Confidence level computation for combining searches
  with small statistics}'',} \textit{ Nucl. Instrum. Meth. A} \textbf{ 434}
  (1999) 435,
  \href{http://dx.doi.org/10.1016/S0168-9002(99)00498-2}{\doi{10.1016/S0168-9002(99)00498-2}}.

\bibitem{LHC-HCG}
\href {http://cdsweb.cern.ch/record/1379837}{{ATLAS and CMS Collaborations, LHC
  Higgs Combination Group}, ``{Procedure for the {LHC} {H}iggs boson search
  combination in Summer 2011}'',} Technical Report ATL-PHYS-PUB 2011-11,
  CMS-NOTE 2011-005, 2011.

\bibitem{GMH}
\hrefCMSnoop {}{{H. Georgi and M. Machacek}, ``Doubly charged Higgs bosons'',}
  \textit{ Nucl. Phys. B} \textbf{ 262} (1985) 463,
  \href{http://dx.doi.org/10.1016/0550-3213(85)90325-6}{\doi{10.1016/0550-3213(85)90325-6}}.

\end{thebibliography}\endgroup
